\documentclass[reprint,prl,twocolumn,showpacs,superscriptaddress,aps,longbibliography]{revtex4-2}

\usepackage[colorlinks=true,citecolor=blue]{hyperref} 
\usepackage{graphicx}
\usepackage{bm}
\usepackage{amsmath}
\usepackage{amssymb}
\usepackage{comment}
\usepackage{braket}
\usepackage{booktabs}
\newcommand{\dIdV}{d$I$/d$V$}
\DeclareGraphicsExtensions{.png,.jpg,.eps}
\usepackage{xcolor}

\begin{document}
\title{Molecular Hamiltonian learning from setpoint-dependent scanning tunneling spectroscopy}

\author{Greta Lupi}
\affiliation{Department of Applied Physics, Aalto University, 02150 Espoo, Finland}
\author{Adolfo O. Fumega}
\affiliation{Department of Applied Physics, Aalto University, 02150 Espoo, Finland}
\author{Mohammad Amini}
\affiliation{Department of Applied Physics, Aalto University, 02150 Espoo, Finland}
\author{Robert Drost}
\affiliation{Department of Applied Physics, Aalto University, 02150 Espoo, Finland}
\author{Peter Liljeroth}
\affiliation{Department of Applied Physics, Aalto University, 02150 Espoo, Finland}
\author{Jose L. Lado}
\affiliation{Department of Applied Physics, Aalto University, 02150 Espoo, Finland}

\date{\today}

\begin{abstract}
Molecular quantum magnets adsorbed on surfaces exhibit rich spin and orbital excitations that can be probed by scanning tunneling microscopy with inelastic electron tunneling spectroscopy (STM-IETS). However, the quantitative extraction of the underlying multiorbital Hamiltonian from experimental spectra remains a fundamental challenge. Here, we introduce molecular Hamiltonian learning, a machine learning strategy that infers the microscopic Hamiltonian parameters of a single adsorbed molecule directly from the setpoint-dependence of STM-IETS data. The method leverages the systematic evolution of spectral features as the STM tip tunes the local electrostatic environment for
different tip-sample distances. We demonstrate this approach on iron phthalocyanine on ferroelectric SnTe, training our
algorithm on theory spectra from a realistic multiorbital model, including spin-orbit coupling, electrostatic interactions, local crystal field, and substrate effects. The algorithm, trained solely on theoretical many-body simulations, allows reconstructing Hamiltonian parameters directly from experimental spectra. Our manuscript establishes a flexible and automated strategy for Hamiltonian reconstruction from STM-IETS, transforming setpoint-dependent spectroscopy into quantitative characterization of quantum materials at the atomic scale.
\end{abstract}

\maketitle

Understanding and controlling spin and orbital excitations in individual molecules adsorbed on surfaces is central to quantum magnetism~\cite{Gatteschi2006, Bogani2008}, molecular spintronics~\cite{Sanvito2011, Urdampilleta2011}, the realization of molecular spin qubits~\cite{Atzori2019, Godfrin2017} and atomic-scale sensing~\cite{Heinrich2004, Hirjibehedin2006}. Scanning tunneling microscopy (STM), combined with inelastic electron tunneling spectroscopy (IETS), offers direct access to these excitations~\cite{Stipe1998, Heinrich2004, Hirjibehedin2006}.
However, quantitatively extracting microscopic Hamiltonian parameters from experimental spectra remains difficult, typically requiring manual peak fitting, heuristic models, and restrictive prior assumptions about the spin state~\cite{Baumann2015, Natterer2017, Campbell2016}.
Molecular excitations are highly sensitive to the local electrostatic environment, which can be systematically tuned during experiments, for example, by varying the STM setpoint current to adjust the tip-molecule distance, thereby modifying hybridization and electric fields~\cite{Ogawa2007,Paschke2025}.
As a result, setpoint-dependence can be exploited as an active strategy to probe nanoscale quantum physics~\cite{Chatzopoulos2021,Fan2021,Li2025,Ko2022,Ko2023,Karan2024}.

\begin{figure}[t]
    \centering
    \includegraphics[width=\linewidth]{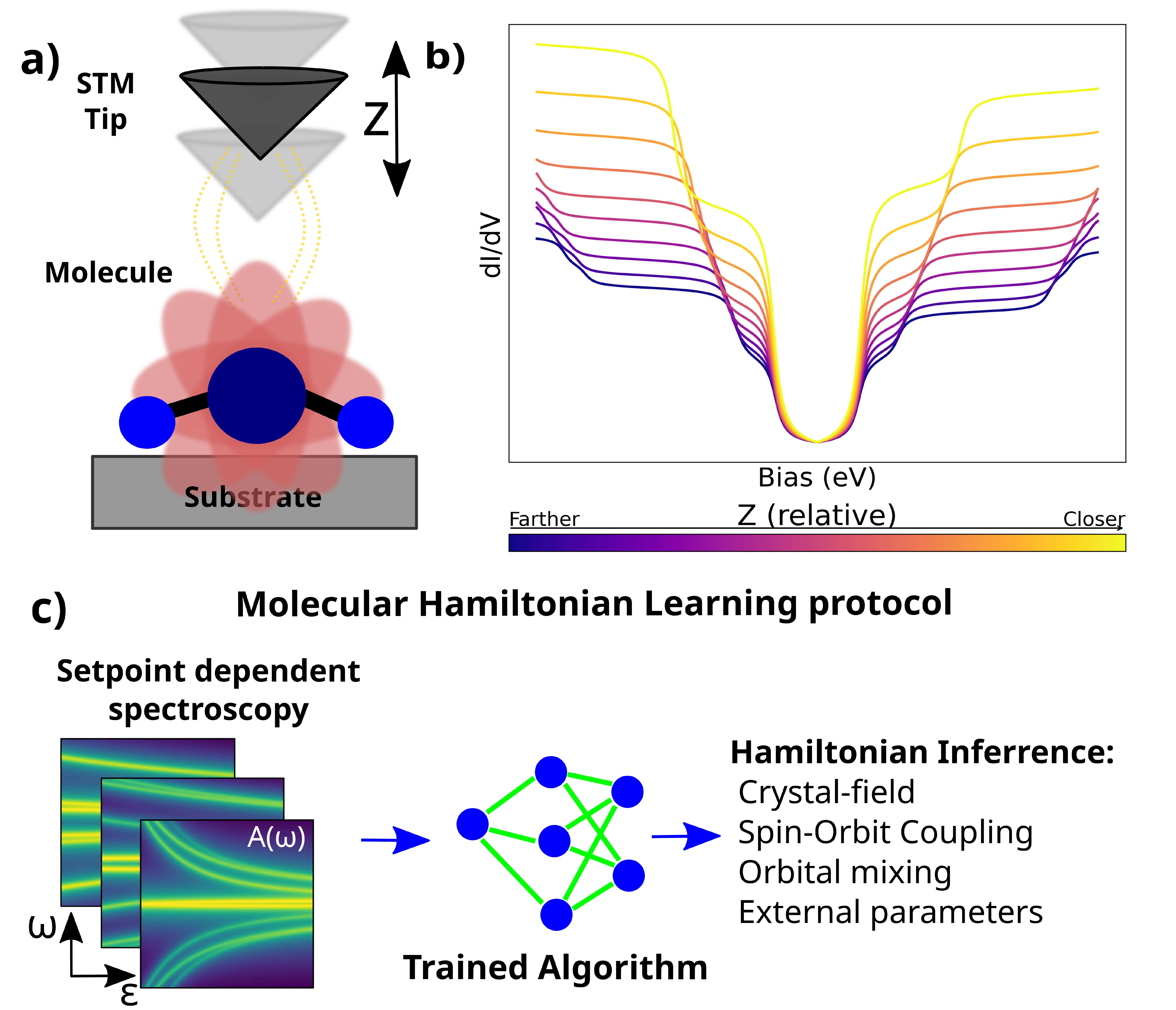}
     \caption{
    (\textbf{a}) STM measurement principle: varying the tip–molecule distance $z$ via the setpoint current tunes the local electrostatic environment and orbital energies.
    (\textbf{b}) Evolution of simulated \dIdV spectra with decreasing $z$
    (\textbf{c}) Scheme of Hamiltonian learning methodology.
    }
    \label{fig:fig1}
\end{figure}

Hamiltonian learning has recently emerged as a powerful paradigm for extracting microscopic models directly from experimental observables in a variety of quantum platforms ~\cite{Wang2017,Garrison2018,Barey2019,Che2021,Gebhart2023,Valenti2022, Gebhart2023,rupp,Simard2025}. This 
approach has been implemented using various machine learning strategies~\cite{PhysRevLett.127.140502,PhysRevLett.130.200403,PhysRevA.110.062421,PhysRevLett.133.040802,Yu2023robustefficient,Dutkiewicz2024advantageofquantum,Haah2022,PRXQuantum.5.040306,Khosravian_2024, Karjalainen2023, vandriel2024, Lupi2025, Karjalainen2025,Koch2022, Koch2023,GarcaEsteban2024}. Among them, supervised learning has enabled the extraction of physical parameters from real-space tomography data ~\cite{Khosravian_2024, Karjalainen2023, vandriel2024, Lupi2025, Karjalainen2025}, while generative models have been employed to learn and predict spectroscopic features ~\cite{Koch2022,Koch2023,GarcaEsteban2024,PantisSimut2023}. 
Among others, Hamiltonian learning with STM
has enabled mapping complex spectroscopic data into Hamiltonian parameters of
quantum magnets~\cite{Koch2025} and correlated two-dimensional materials~\cite{Sobral2023}. 
While the tip-sample distance is commonly taken
as a fixed parameter, setpoint-dependence
spectroscopy can provide unique information
for Hamiltonian learning with STM.

Here, we introduce a molecular Hamiltonian learning strategy
for STM-IETS
that leverages setpoint-dependent spectroscopy.
In our approach, we vary the tip–sample distance to controllably modify the electrostatic environment of the molecule and use STM-IETS to read out the resulting changes in the many-body excitation spectrum (Fig.~\ref{fig:fig1}a–c).
We demonstrate this method with STM-IETS measurements of
iron phthalocyanine (FePc) molecules on top of SnTe, and show that our
approach, trained in quantum many-body simulations, enables extracting key quantities, such as crystal field splitting, spin–orbit coupling, orbital energies,
and substrate-induced orbital mixing.
Importantly, this methodology is
trained exclusively in computational data, and 
enables
performing Hamiltonian learning directly from experimental measurements in individual molecules.

\begin{figure}[t]
    \centering
    \includegraphics[width=0.77\linewidth]{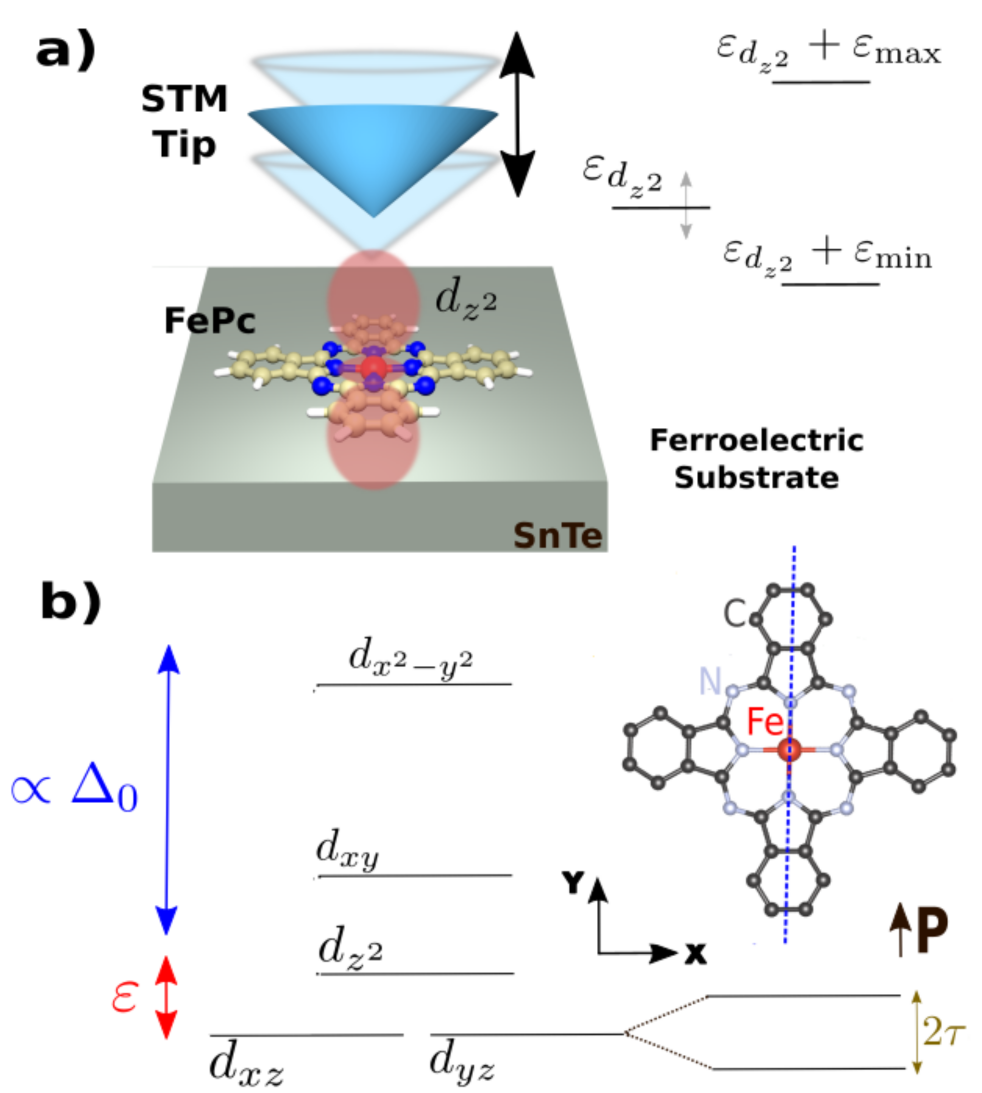}
    \caption{
    (\textbf{a}) Schematic side view: a FePc molecule adsorbed on SnTe.
    (\textbf{b}) Top view of FePc and crystal field splitting of the $3d$ orbitals. The ferroelectric substrate polarization $\mathbf{P}$ induces a rectangular distortion of the ligand-field symmetry. 
    }
    \label{fig:fig2}
\end{figure}

A single magnetic molecule adsorbed on a surface can be described by an effective multiorbital Hamiltonian acting on its open shell  orbitals, in particular d-orbitals for transition-metal-based molecules~\cite{Gatteschi2006,Malavolti2018,PhysRevLett.131.086701,Kezilebieke2019,Wang2021,Lin2023}, molecular p-orbitals in nanographenes~\cite{Mishra2019,Mishra2021,Mishra2021n,PhysRevLett.124.177201},
and f-orbitals for rare-earth systems~\cite{Reale2024,Donati2021,Singha2021}. While the precise orbital composition depends on the molecular symmetry and ligand environment, the excitation spectrum accessible by STM-IETS is generically governed by the inter-orbital hybridizations, often referred to as crystal field splitting,
the many-body Coulomb interactions, and spin–orbit coupling~\cite{Hirjibehedin2006,Heinrich2004}. 
Generally, the system can be described by a multiorbital Hamiltonian of the form
\begin{equation}
    \mathcal{H} = \mathcal{H}_{\text{CF}} (z) + \mathcal{H}_{\text{Coulomb}} + \mathcal{H}_{\text{SOC}},
\end{equation}
where $\mathcal{H}_{\text{CF}}(z)$ is the crystal field Hamiltonian, $\mathcal{H}_{\text{Coulomb}}$ describes the many-body Coulomb interaction, and $\mathcal{H}_{\text{SOC}}$ is the spin-orbit coupling.

The crystal field Hamiltonian $\mathcal{H}_{\text{CF}}(z)$ can be decomposed into two components:
\begin{equation}
    \mathcal{H}_{\text{CF}} (z) = \mathcal{H}^{0}_{\text{CF}} + \mathcal{H}^{\text{tip}}_{\text{CF}} (z).
\end{equation}
The first term, $\mathcal{H}^{0}_{\text{CF}}$, corresponds to the setpoint-independent crystal field, which stems
from the geometry of the molecule hosting
the open-shell correlated electrons 
and the presence of a static substrate~\cite{Atanasov2012}. The second term, $\mathcal{H}^{\text{tip}}_{\text{CF}}(z)$, stems from the proximity of the STM tip and is tunable by the tip–molecule distance $z$~\cite{Natterer2017}.
The crystal field terms take the form:
\begin{align}
    \mathcal{H}^{0}_{\text{CF}} &= \sum_{i,j, s} t_{ij} \, c^\dagger_{i,s} c_{j,s}, \label{eq:H0_CF} \\
    \mathcal{H}^{\text{tip}}_{\text{CF}} (z) &= \sum_{i,j, s} \gamma_{ij}(z) \, c^\dagger_{i,s} c_{j,s}, \label{eq:Htip_CF}
\end{align}
where $c^\dagger_{i,s}$ ($c_{i,s}$) creates (annihilates) an electron in orbital $i$ with spin $s$, $t_{ij}$ parameterizes the intrinsic orbital energies ($i=j$) and inter-orbital hybridizations ($i \neq j$), and $\gamma_{ij}(z)$ encodes the tip-induced perturbation to the crystal field.

The many-body Coulomb interaction is given by
\begin{equation}
    \mathcal{H}_{\text{Coulomb}} = \sum_{ijkls s'} V_{ijkl} \, c^\dagger_{i,s} c^\dagger_{k,s'} c_{l,s'} c_{j,s}, \label{eq:Coulomb}
\end{equation}
where $V_{ijkl}$ are the matrix elements of the Coulomb interaction in the molecule. 

The spin–orbit coupling Hamiltonian takes the form
\begin{equation}
    \mathcal{H}_{\text{SOC}} = \lambda_{\text{SOC}} \sum_{ij\alpha s s'}\ell^\alpha_{ij} \, \sigma^\alpha_{ss'} \, c^\dagger_{i,s} c_{j,s'}, \label{eq:SOC}
\end{equation}
where $\lambda_{\text{SOC}}$ is the SOC coupling strength, $\boldsymbol{\ell} = (\ell^x,\ell^y,\ell^z)$ is the angular momentum operator
in the orbital basis, and $\boldsymbol{\sigma} = (\sigma^x,\sigma^y,\sigma^z)$ are the spin Pauli matrices.

The Hamiltonian above allows accounting for the low-energy physics of diverse transition-metal complexes
such as phthalocyanines, which in our case will correspond to FePc. 
The different Hamiltonian parameters directly determine the hierarchy of spin and orbital excitations, which appear in STM-IETS as characteristic inelastic steps, and subsequently as peaks in the d$^2 I$/d$V^2$. Inelastic spin-flip excitations are primarily governed by the interplay between crystal field splitting and spin–orbit coupling, appearing at energies bounded by the
SOC of the 3d element of the Pc, which for FePc corresponds to a few tens of meV. 
On the other hand, inelastic orbital excitations are processes in which the incoming and outgoing electrons
come from different orbitals. 
In the case of highly symmetric environments with close to orbital degeneracy, they can give rise to excitations at energies comparable to spin-flip transitions. The STM–IETS conductance spectra \dIdV correspond to the bias integral of the quantity we denote as $d^2I/dV^2$, which is proportional to the dynamical correlator $A(\omega)$. The \dIdV is used as input to our Hamiltonian learning model.
Inelastic step positions and their evolution with tip–molecule coupling encode the Hamiltonian parameters and can be computed from the many-body Hamiltonian.

For FePc, the open shell orbitals correspond to the different Wannier-like orbitals stemming
from the d manifold of Fe. Due to the symmetry of the orbitals, the orbital that couples the
strongest to an STM tip is the Wannier $d_{z^2}$ orbital, which in the following we will take as the orbital
that mediates the tunneling between the molecule and the STM tip.
The spin-flip orbital-conserving contribution to d$^2 I$/d$V^2$ takes the form
\begin{equation}
    A_{\text{flip}}(\omega) = \sum_{\nu} \bra{\Omega} \hat{S}^{\nu}_{d_{z^2}} \, \delta(\omega - \mathcal{H} + E_{\Omega}) \, \hat{S}^{\nu}_{d_{z^2}} \ket{\Omega}, \label{eq:S_flip}
\end{equation}
where $\ket{\Omega}$ is the many-body ground state with energy $E_{\Omega}$, and $\hat{S}^{\alpha}_{d_{z^2}}$ is the $\alpha$-component of the spin operator projected onto the $d_{z^2}$ orbital~\cite{Heinrich2004}. This inelastic contribution to the current
from our many-body fermionic Hamiltonian
is analogous to the spin-flip excitation
that can be captured with a purely spin model~\cite{Ternes2015}.

The orbital cotunneling contribution accounts for processes where the incoming and outgoing electrons stem from different
orbitals. This contribution to inelastic spectroscopy, including both spin-conserving and spin-flip inelastic processes, takes the form
\begin{equation}
    A_{\text{orb}}(\omega) = \sum_{\alpha,s,s'} \Theta(\omega) \, C^{s,s'}_{d_{z^2},\alpha}(\omega) + [1-\Theta(\omega)] \, C^{s,s'}_{\alpha,d_{z^2}}(\omega), \label{eq:A_orb}
\end{equation}
where $\Theta(\omega) = \frac{1}{2} \left ( 1 + \text{sign}(\omega) \right )$ is the Heaviside step function, and
$C^{s,s'}_{d_{z^2},\alpha }$ and $C^{s,s'}_{\alpha,d_{z^2}}(\omega)$ are the orbital-switching
inelastic processes for positive and negative bias voltage, which take the form

\begin{equation}
    C^{s,s'}_{\alpha,\beta}(\omega) = \bra{\Omega} c_{\beta,s}^\dagger c_{\alpha,s'} \, \delta(\omega - \mathcal{H} + E_{\Omega}) \, c_{\alpha,s'}^\dagger c_{\beta,s} \ket{\Omega}. 
    \label{eq:Cup}
\end{equation}
It is worth noting that in general $C^{s,s'}_{\alpha,\beta} \ne C^{s,s'}_{\beta,\alpha}$ for $\beta \ne \alpha$, which gives rise to an asymmetry in the height of the
inelastic spectroscopy between positive and negative bias voltages. This must be contrasted with the spin flip contribution $A_{\text{flip}}$,
which by definition is symmetric between positive and negative bias, in the absence of spin-polarized tunneling.
The previous dynamical correlators can be computed by solving the quantum many-body Hamiltonian in the full many-body space.
As the Hamiltonian depends on a variety of parameters, including the crystal field
and spin-orbit coupling, the inelastic spectroscopy will inherit those dependencies, which will be leveraged to perform Hamiltonian learning.

We cast the Hamiltonian inference as a supervised inverse problem: a neural network is trained on theory spectra generated over physically realistic ranges of Hamiltonian parameters. Each spectrum is treated as a 2D image with axes corresponding to bias and a tuning parameter controlled experimentally,
and specifically the setpoint current. The trained algorithm\footnote{Details on dataset generation, preprocessing, and training are provided in the Supplemental Material.}
thus allows mapping the evolution of the spectral fingerprint to the underlying Hamiltonian parameters. The complete strategy is illustrated in Fig.~\ref{fig:fig1}c.

We apply this strategy to a concrete case study: a single Fe-phthalocyanine (FePc) molecule adsorbed on a ferroelectric SnTe substrate~\cite{Amini2023, Amini2025Arxiv}. 
FePc hosts a central Fe$^{2+}$ ion in a nearly square-planar $D_{4h}$ ligand field, which splits the $3d$ shell into $a_{1g}(d_{z^2})$, $b_{1g}(d_{x^2-y^2})$, $b_{2g}(d_{xy})$, and the doubly degenerate $e_g(d_{xz},d_{yz})$ orbitals (as shown in Fig.~\ref{fig:fig2}b)~\cite{Miedema2009}. 
The competition between the crystal field splitting $\Delta_0$ and intra-atomic interactions leads to two possible states $S=1$ and $S=2$.
In the free-standing limit for Fe-Pc, the ground state
has $S=1$, with the $S=2$ manifold lying 
$\sim$200–300~meV higher in energy~\cite{FernandezRodriguez2015}. This
energetic near-degeneracy makes FePc substantially
sensitive to its environment. In particular,
when deposited on SnTe~\cite{Amini2025Arxiv}, FePc stabilizes in the $S=2$ ground state, which is the regime studied here.

When FePc is placed on ferroelectric SnTe (Fig.~\ref{fig:fig2}a), the local symmetry is further reduced. As shown in Fig.~\ref{fig:fig2}b, the degeneracy of $d_{xz}$/$d_{yz}$ is lifted and the two orbitals hybridize, captured by parameter $\tau$ that quantifies the magnitude
of the symmetry breaking~\cite{Amini2023,Chang2020}. Simultaneously, the STM setpoint current controls the vertical tip–molecule distance and produces a tunable Stark shift of the $a_{1g}(d_{z^2})$ on-site energy, which we denote by $\varepsilon$~\cite{Amini2023}. 
From the point of view of the Hamiltonian in Eq. (\ref{eq:Htip_CF}), this gives rise to a tip-dependent perturbation of the 
form $\mathcal{H}^{\text{tip}}_{\text{CF}} (z) = \varepsilon (z) \sum_s c^\dagger_{d_{z^2,s}} c^\dagger_{d_{z^2,s}}$.
The systematic evolution of the excitation spectrum with $\varepsilon (z) \equiv \varepsilon$ serves as the spectroscopic fingerprint for Hamiltonian learning.
The resulting crystal field Hamiltonian $\mathcal{H}_{CF}$ for the Fe $3d$ shell is parameterized by these orbital energies and 
substrate-induced inter-orbital mixing\footnote{Explicit form of the crystal field Hamiltonian is shown in the Supplemental Material}.

A practical challenge is that a mapping between the orbital shift $\varepsilon$ and the tip position, or setpoint-current, 
is not quantitatively known. To address this, we randomly sample the bounds $(\varepsilon_{\min},\varepsilon_{\max})$ during training. Consequently, the ML model learns both the Hamiltonian parameters
controlling the crystal field and spin-orbit coupling,
and the effective Stark-shift window $(\varepsilon_{\text{min}},\varepsilon_{\text{max}})$, 
directly from changes in the inelastic spectroscopy, without assuming a fixed mapping between
tip height and orbital energy.

\begin{figure}[t]
    \centering
    \includegraphics[width=\linewidth]{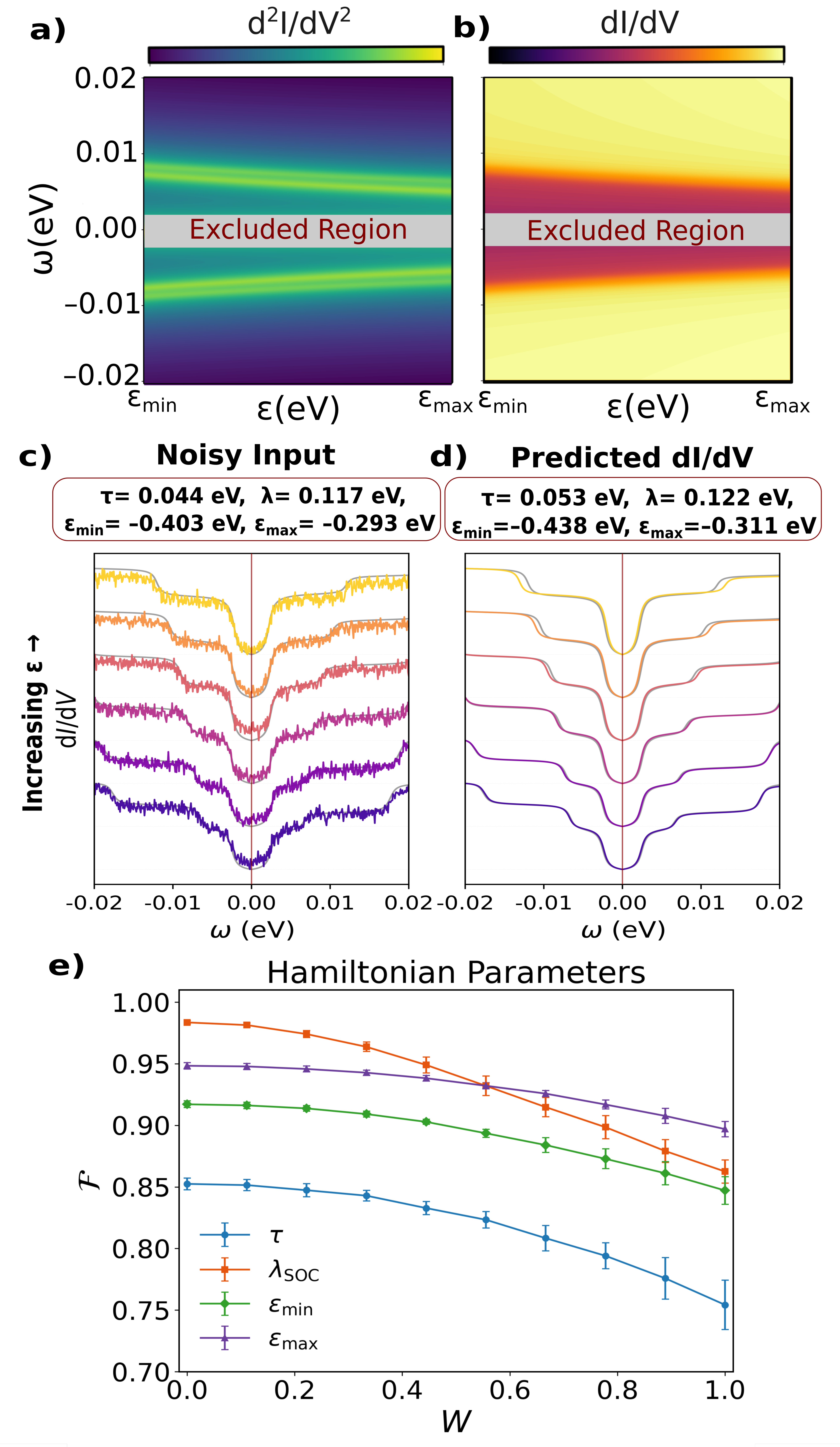}
    \caption{Representative theoretical spectrum: d$^2I$/d$V^2$ (\textbf{a}) and its cumulative integral \dIdV (\textbf{b}), with the central bias window ($\pm 3$ meV) set to zero. The integrated dynamical correlator serves as input to the ML model.
    (\textbf{c}) Test set \dIdV spectrum (gray) with added Gaussian noise ($W=5\%$, colored) used as network input.
    (\textbf{d}) Reconstructed \dIdV obtained from the Hamiltonian parameters predicted by the network for the noisy input in  (\textbf{c}); predicted values are indicated.
    (\textbf{e}) The fidelity $\mathcal{F}$ for Hamiltonian parameters ($\tau$, $\lambda_{\mathrm{SOC}}$, $\varepsilon_{\mathrm{min}}$, $\varepsilon_{\mathrm{max}}$)
    as a function of noise $W$.
    }
    \label{fig:fig3}
\end{figure}

For the FePc molecule in the high-spin regime, $\Delta_0$ has very limited influence on the spectroscopy,
the specific low-energy spectral features are dominated by $\lambda_{\mathrm{SOC}}$, $\tau$, and the Stark-shift window $[\varepsilon_{\min},\varepsilon_{\max}]$ \footnote{See the SI for a detailed discussion of the full crystal field}. A representative sample from our theory dataset is shown in Fig.~\ref{fig:fig3}a-b, displaying both the d$^2I$/d$V^2$ and the corresponding \dIdV. We exclude the central bias region within $\pm 3$ meV, as our theoretical methodology would not capture zero bias anomalies associated with
Kondo physics potentially
arising from sizable coupling with the substrate.


The performance of our Hamiltonian learning algorithm is first evaluated by testing its reconstruction fidelity under progressively increasing Gaussian noise. 
Noise is included by adding zero-mean Gaussian noise with standard deviation $W$ as
\begin{equation}
dI/dV_{\text{noisy}}(\varepsilon,\omega) = dI/dV_{\text{test}}(\varepsilon,\omega) + \chi_{\text{noise}}(\varepsilon,\omega),
\end{equation}
where $\chi_{\text{noise}}(\varepsilon,\omega)$ is a random variable sampled from a Gaussian distribution with variance $W^2$, satisfying $\langle \chi_{\text{noise}}(\varepsilon,\omega) \rangle = 0$ and $\langle \chi_{\text{noise}}^2(\varepsilon,\omega) \rangle = W^2$. An example of a spectrum corrupted with $W = 0.05$ is shown in Fig.~\ref{fig:fig3}c. This noisy \dIdV map is then fed into the trained neural network, which outputs a set of predicted Hamiltonian parameters. Using these predicted parameters, we recalculate the corresponding \dIdV spectrum via our many-body solver. As demonstrated in Fig.~\ref{fig:fig3}d, the network successfully reconstructs a spectrum that closely matches the original, noiseless one while accurately inferring the underlying microscopic parameters (listed in the figure panel).
We then quantify this performance statistically across the entire test set for increasing values of $W$ by measuring the fidelity $\mathcal{F}_\Lambda$ of each one of the Hamiltonian parameters $\Lambda=(\tau,\lambda_{SOC},\varepsilon_{min},\varepsilon_{max})$, defined as ~\cite{Khosravian_2024,Lupi2025,LIU2022106426, rupp}
\begin{equation}
{\cal{F}}_\Lambda 
=
\frac{\left|\langle\Lambda^{\mathrm{pred}}\Lambda^{\mathrm{true}}\rangle-\langle\Lambda^{\mathrm{pred}}\rangle\langle\Lambda^{\mathrm{true}}\rangle \right|}{\sqrt{\langle \langle \Lambda^{\mathrm{pred}} \rangle\rangle 
\langle \langle \Lambda^{\mathrm{true}} \rangle\rangle}}
\label{fidelity}
\end{equation}
where $\langle \langle \Lambda^{\mathrm{pred}} \rangle\rangle =  \langle(\Lambda^{\mathrm{pred}})^{2}\rangle-\langle\Lambda^{\mathrm{pred}}\rangle^{2} $
and
$ \langle \langle \Lambda^{\mathrm{true}} \rangle\rangle  = \langle(\Lambda^{\mathrm{true}})^{2}\rangle-\langle\Lambda^{\mathrm{true}}\rangle^{2}$
are cumulants of the predicted and real parameters.
A value of $\mathcal{F}=1$ indicates perfect reconstruction,
whereas a value $\mathcal{F}=0$ corresponds to no predictive ability.
As shown in Fig. \ref{fig:fig3}e, the fidelity
all parameters, $\mathcal{F}$ remains close to or above $0.9$ at low noise levels, confirming the model's intrinsic accuracy on clean theory data. Crucially, $\mathcal{F}$ exhibits a gradual, monotonic decline as the noise standard deviation $W$ approaches $100\%$ of the typical signal amplitude. This controlled degradation indicates that the network learns physically meaningful spectral features rather than noise.
Parameters governing direct spectral shifts, such as the Stark-shift bounds and the SOC strength show the highest resilience, while the ferroelectric strength is more sensitive, as its related to finer spectral details.

\begin{figure}[t]
    \centering
    \includegraphics[width=\linewidth]{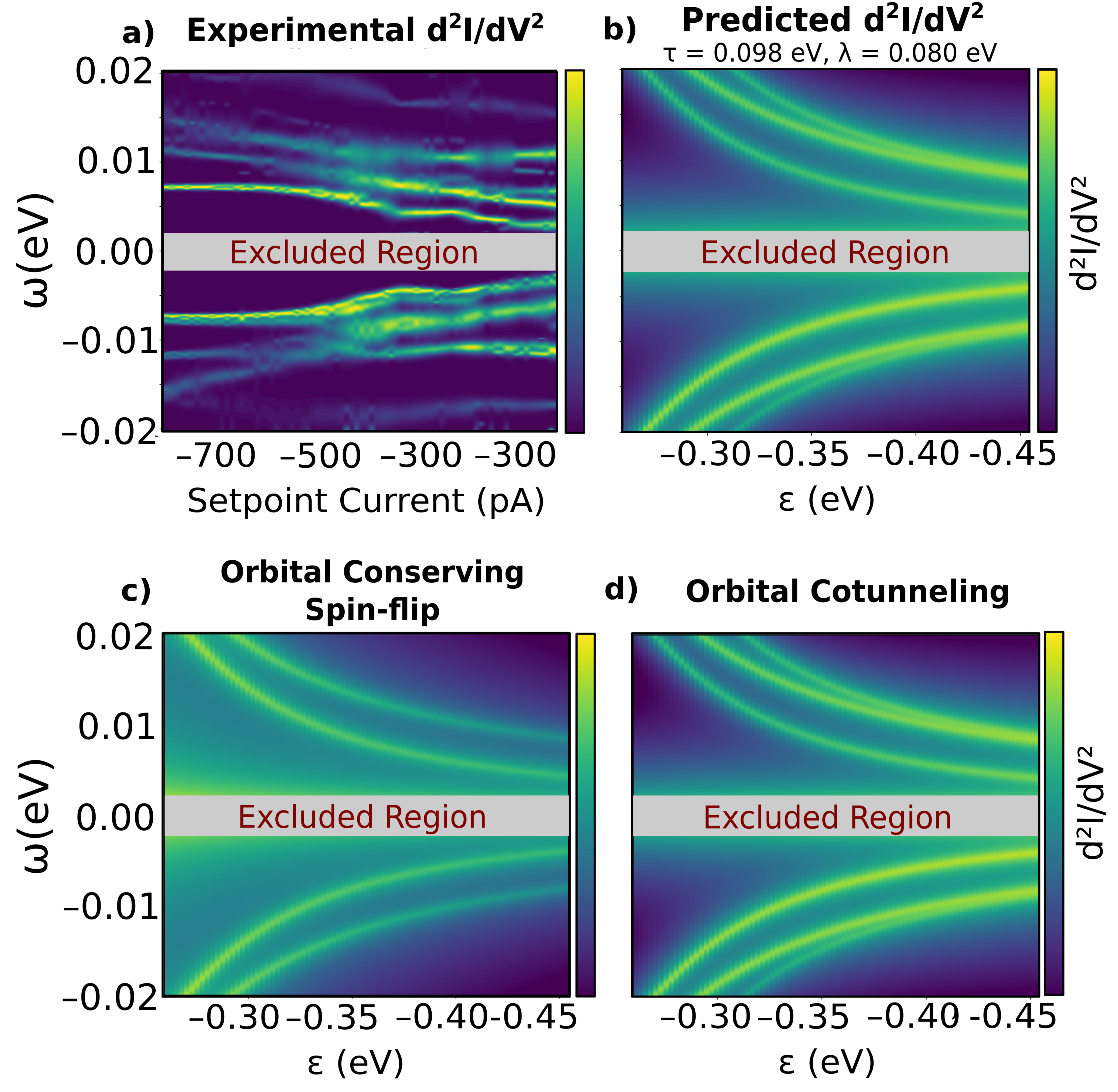}
    \caption{
  (\textbf{a}) Experimental $d^2I/dV^2$ spectrum of a single FePc molecule on SnTe. The derivative is obtained numerically from the measured \dIdV signal. (\textbf{b}) Reconstructed dynamical correlator from the Hamiltonian parameters predicted by the model from the input in (\textbf{a}). (\textbf{c}) Orbital-conserving spin-flip and (\textbf{d}) orbital cotunneling channel contributions to the total reconstructed excitation spectrum.}
    \label{fig:fig4}
\end{figure}

We apply our trained Hamiltonian learning model directly to the preprocessed experimental spectra (Fig.~\ref{fig:fig4}a). Prior to inference, the experimental \dIdV curves are processed identically to the theory training set: they are normalized, and the central bias region within $\pm 3$ meV is excluded. 
One important question is how the
setpoint is mapped to the change in orbital energy.
The tunneling current depends quadratically on the overlap between molecular and tip orbitals, as does the second-order perturbative correction to the $d_{z^2}$ orbital energy; both quantities depend exponentially on the tip–molecule distance.
As a result, we would obtain that $\varepsilon$ and $I$
follow a linear relationship $\varepsilon = c_1 I + c_0$. 
Thus, for the experimental data, varying the setpoint current provides a monotonic tuning of $\varepsilon$ over a certain interval~\cite{Tersoff1985, Limot2003, Natterer2017}. It is finally worth noting that setpoint-dependent mechanical
deformations of the molecule would provide a non-linear
correction to the $I-\varepsilon$ mapping.

The ML model allows us to invert these experimental inputs, yielding the reconstructed $d^2I/dV^2$ shown in Fig.~\ref{fig:fig4}b and the predicted microscopic parameters. The ML model extracts a symmetry-breaking field strength of $\tau = 0.098$ eV and a spin–orbit coupling constant of $\lambda_{\mathrm{SOC}} = 0.080$ eV from the experimental spectra. The magnitude of $\tau$ is consistent with the scale of band bending expected at the ferroelectric SnTe surface ($2\tau  \approx 0.2$ eV)~\cite{Chang2016}. Interestingly, the extracted $\lambda_{\mathrm{SOC}}$ is nearly a factor of two
larger than the expected
intrinsic values for isolated
FePc molecules~\cite{Kroll2009,Natoli2018}. We attribute this substantial 
enhancement to the interfacial electronic coupling with the SnTe substrate~\cite{Liang2017,Shen2024,Lv2025}. Heavy elements like Sn and Te possess strong intrinsic spin–orbit coupling, which can be imprinted onto adsorbed molecules through wavefunction hybridization, a known proximity effect in van der Waals heterostructures~\cite{Avsar2014,Tu2022,Choi2023,Masseroni2024}.

Finally, it is worth disentangling the pure spin flip (Fig.~\ref{fig:fig4}c),
and orbital cotunneling contributions (Fig.~\ref{fig:fig4}d).
We observe that the spin-flip contribution
Fig.~\ref{fig:fig4}c shows two excitation branches,
whereas the orbital cotunneling contribution
shows three branches Fig.~\ref{fig:fig4}d.
As the experiment (Fig.~\ref{fig:fig4}a) features three
excitation branches, the previous disentanglement
shows that a branch of many-body
modes comes from orbital cotunneling. Specifically,
orbital cotunneling inherently emerges from a many-body
fermionic model, a feature not directly
accountable by a purely spin model for FePc.

Our calculations rely on a minimal model for the IETS of the system, while more complete descriptions are possible.
First, our crystal
field does not consider hybridization
between $d_{z^2}$ and the other orbitals, which could be present
experimentally due to the substrate~\cite{Amini2025Arxiv}. 
Second, the inelastic contributions
assume that tunneling happens through
a tip s-orbital that only hybridizes with the $d_{z^2}$
Wannier orbital, whereas for very close distances
other orbital contributions from the tip may be present.
The two points above could be accounted for by extracting
a multiorbital Hamiltonian accounting for both molecule, tip
and sample using Wannierization and first principles calculations~\cite{RevModPhys.84.1419,Thygesen2005}.
Third, our calculations assume that inelastic channel
alone accounts for the spectroscopy, whereas additional
channels, involving Kondo or inelastic excitations from the substrate, can lead to sizable changes in
the spectroscopy~\cite{Ternes2015}. 
And finally, at very short distances, current-driven
non-equilibrium effects can impact the state of the molecule~\cite{Loth2010} and dominate the change of the spectroscopy.
While the previous effects
are not considered in our calculations, 
their impact could be systematically included within our Hamiltonian learning strategy, potentially improving the agreement of the
results shown in Fig.~\ref{fig:fig4}. 

To summarize, we have presented a 
molecular Hamiltonian learning strategy
to quantitatively infer the microscopic Hamiltonian of a single
molecule directly from STM-IETS. 
Our methodology exploits the dependence of the spectroscopy on the tip-sample distance,
a sometimes undesirable feature, turning it into a tool to learn subtle interactions in nanoscale quantum many-body systems.
Our algorithm is trained on quantum many-body calculations, 
and learns the nonlinear map between complex spectral fingerprints and the underlying crystal field, spin–orbit, and
substrate-induced symmetry-breaking parameters.
We demonstrated this strategy experimentally on FePc on ferroelectric SnTe, enabling us to 
extract its microscopic parameters, including substrate-enhanced spin–orbit coupling and symmetry-breaking field strength. 
Beyond SnTe, our approach can be extended to map interfacial screening, chemical gating, and substrate coupling effects, positioning STM-enabled Hamiltonian learning as a versatile route to uncover microscopic quantum behavior at the single-molecule limit.

\textbf{Data Availability:} 
Both the source code~\cite{lupi_molecular_hamiltonian_2025} and the datasets~\cite{lupi_dataset_2025} are publicly available.

\textbf{Acknowledgments:}
We acknowledge financial support from the European Research Council (ERC-2023-AdG GETREAL No.~101142364, ERC-2024-CoG ULTRATWISTROICS No.~101170477), the Research Council of Finland Projects Nos. 370912, 369367, 368478, and 353839, InstituteQ, the Jane and Aatos Erkko Foundation, the Finnish Centre of Excellence in Quantum Materials QMAT (project No. 374166) and the Finnish Quantum Flagship (project No.~358877). We acknowledge the computational resources provided by the Aalto Science-IT project. We thank R. Koch and A. Cahlik for useful discussions.
\bibliography{ref}

\end{document}